\documentclass[aps,prl,twocolumn,10 pt,superscriptaddress,showpacs]{revtex4}
\usepackage{amssymb}
\usepackage{graphicx}
\begin{document}
\title{Structure of the flux lines lattice in $NbSe_2$: equilibrium state and influence of the magnetic history.}
\author{A. Pautrat}
\affiliation{Laboratoire CRISMAT, UMR 6508 du CNRS, ENSICAEN et Universit$\acute{e}$ de Caen, 6 Bd Mar$\acute{e}$chal Juin, F-14050 Caen 4, France.} 
\author{M. Aburas}
\affiliation{Laboratoire CRISMAT, UMR 6508 du CNRS, ENSICAEN et Universit$\acute{e}$ de Caen, 6 Bd Mar$\acute{e}$chal Juin, F-14050 Caen 4, France.} 
\author{Ch. Simon}
\affiliation{Laboratoire CRISMAT, UMR 6508 du CNRS, ENSICAEN et Universit$\acute{e}$ de Caen, 6 Bd Mar$\acute{e}$chal Juin, F-14050 Caen 4, France.} 
\author{P. Mathieu}
\affiliation{Laboratoire Pierre Aigrain de l'Ecole Normale Sup\'erieure, UMR 8551 du
CNRS, associ\'ee aux universit\'es Paris 6 et 7, 75231 Paris Cedex5, France.} 
\author{A. Br\^{u}let}
\affiliation{Laboratoire L\'{e}on Brillouin, \\ CEA Saclay, 91191
Gif/Yvette, France.}
\author{C. D. Dewhurst}
\affiliation{Institut Laue Langevin, 6 rue Jules Horowitz,
Grenoble, France.}
\author{S. Bhattacharya}
\affiliation{Tata Institute of Fundamental Research, Mumbai-400005, India.}        
\author{M. J. Higgins}
\affiliation{Department of Physics, Rutgers University, Piscataway, New Jersey 08855, USA.}

\begin{abstract}
 We have performed small-angle neutron scattering (SANS) of the flux line lattice (FLL) in a $Fe$ doped $NbSe_2$ sample which presents
 a large peak effect in the critical current. The scattered intensity and the width of the Bragg peaks of the equilibrium
 FLL indicate an ordered structure in the peak effect zone.
The history dependence in the FLL structure has been studied using field cooled and zero field cooled procedures,
 and each state shows the same intensity of Bragg scattering and good orientational order.
 These results strongly suggest that the peak effect is unrelated to a bulk disordering transition,
and confirm the role of a heterogeneous distribution of screening current.   
\end{abstract}

\newpage
\maketitle

The peak effect in the critical current of a type II superconductor is a long standing problem in the field of superconductivity.
 Since its observation \cite{leblanc,berlincourt}, no clear consensus has emerged as to its origin.
 Most models are based on the Pippard idea \cite{pippard} who attributed the peak effect
 to the loss of rigidity of the flux lines lattice (FLL), which falls close to $B_{c2}$ more rapidly than the
 pinning strength due to inhomogeneities. Larkin and Ovchinikov (LO) have treated the complex problem of the elastic FLL
 in the presence of random pinning centers \cite{larkin}. The main issues are the statistical calculations of a volume $V_c$ over which the
 lattice remains correlated and the corresponding pinning force per unit volume $F_p \propto V_c^{-1/2}$.
 In addition to the intense theoretical interest,
there is the possible link between the FLL bulk disorder and the critical current. Other interpretations of the critical current deal with dominant pinning of flux lines by the surface \cite{patrice,niob}, without significant role of FLL bulk order.
Recently, the explanation of the peak effect
 as a genuine phase transition has been addressed both theoretically and experimentally, and the peak effect is now considered
as a classical example of an order-disorder transition in an elastic system \cite{giam}. A superconductor with a peak effect is
 remarkable by its anomalous transport properties. They have been tackled by quite thorough series of measurements in $NbSe_2$,
 with a large focus on the metastable properties \cite{transport}.
Two coexisting macroscopic FLL states with different
 pinning strengths were observed \cite{shobo} and the kinetics between these two states can explain most of the anomalous transport properties.
  Paltiel et al have also shown that the critical current
 in the peak effect zone is heterogeneous and more important close to the edge of the samples \cite{paltiel}. However, the technique of Hall probe
 does not allow concluding on the structure of the FLL. The genuine nature of FLL states with the two different critical currents remains unknown, even if one is usually assumed to be much more disordered following LO approach \cite{paltiel}.
 However, this latter assumption needs to be experimentally controlled using a structural probe: indeed, a correlation between the critical current and the FLL bulk order can not be $\textit{a priori}$ assumed,
as shown by different Small Angle Neutron Scattering (SANS) experiments \cite{thorel,bisco,paxy}.
 When no change in FLL order is observed whereas the critical current is strongly modified \cite{bisco},
 an interpretation could be that the pertinent structural distortions are at a too small scale to be resolved.
 However, when important changes in the FLL order are seen without any modification of the critical current \cite{thorel, paxy},
 it shows more clearly that the related FLL order is not pertinent for the critical current and the dominant pinning mechanism was shown to be at the surface.
SANS in the presence of a peak effect has been mostly used in $Nb$. Measurements show differences in the FLL structure 
 after field cooling and zero-field cooling and it was proposed to arise from superheating and supercooling of metastable states
 as expected for a genuine first order (melting) transition  \cite{ling,danii}. Other authors proposed that the results would be
 better understood involving a crossover into a disordered state \cite{comment}.
 The absence of intrisinc (thermally driven) melting transition in $Nb$ is supported by the persistence of the Abrikosov lattice up to $B_{c2}$ in a purest sample which does not present the peak effect \cite{tedniobium}. 
$Fe$ doped $NbSe_2$ systematically displays a marked
 and extended peak effect, and then was studied as a school case using transport measurements \cite{transport}. In addition, no correlation was found between
 the topology of the FLL structure and the enhancement of the critical current in $NbSe_2$ using the decoration technique \cite{fasano}.
 This makes this sample particularly interesting for a SANS study of equilibrium and metastable FLL, in order to clarify their nature and the link with the pinning properties.

 Our sample is a large single crystal of $2H-NbSe_2$ doped with 100 ppm of $Fe$ ($T_c=5.80 K$), which was used in \cite{paxy}. 
SANS experiments were performed at ILL (Grenoble-France) using D22 and D11 spectrometers.
 The neutron beam was parallel to the magnetic field and no external current was applied. In this geometry, the higher angular resolution is obtained on the rocking curves which measure correlation along the flux lines.
 The neutron wavelength was $\lambda = 8$ and $10$ $\AA$ with
 a spread $\Delta \lambda /\lambda \approx 10 \%$. Depending on the diffraction vector $Q$ of the FLL Bragg planes and
 in order to optimize the intensity, different collimation and sample-detector distances were used. The beam was collimated in the middle of the sample.
 Background scattering, obtained from the sample in the normal state ($B>B_{c2}$) can be subtracted to reveal the 2D FLL image (fig.1).
 All the measurements were made at a temperature $T=2K$, using field cooling (FC: magnetic field applied at $T>T_c$)
 or zero field cooling (ZFC: magnetic field applied at $T=2K$) procedures.

A crucial point for a quantitative interpretation of the results is the value of the upper critical field $B_{c2}$. 
It was measured using different techniques (magnetization, resistivity, $V(I)$ curves) on a small crystal cleaved from the large one.
 $V(I)$ curves were also performed on the whole crystal used for SANS measurements.
 All measurements show a sharp increase of the critical current after ZFC at $B=1.4 T$ and a second critical field $B_{c2}\approx 2.55 T$ at $T=2K$,  as shown in fig.1 (ZFC procedure avoids much of metastability when a peak effect is present).
We observe also strong evidences of surface superconductivity for $B>B_{c2}$. One can estimate the field where the peak effect occurs as the midpoint of the abrupt increase of $I_c(B)$.
 This gives $B_{peak}\approx 1.5T$, i.e. $B_{peak}/B_{c2}\approx 0.6$ in agreement with previous reports \cite{anderson}.
First series of SANS measurements relate to the structure of FLL in the area of the peak effect, to see if there exists a transition toward a disordered phase.
Such measurements were made for applied magnetic fields from $B=0.4T$ to $2T$, using the ZFC procedure to directly compare with the critical current measurements shown in fig.1.
 For fields larger than $B>2T$, the scattered intensity
from the FLL became too small to measure with counting times of 4 Hours on $D22$, but this value is however much higher than $B_{peak}$.
 Such measurements always show a hexagonal
 lattice with sharp Bragg peaks and a diffraction vector $Q$ corresponding to the applied magnetic field.
For each field value, we have measured the full rocking curve to get the scattered intensity of the (1,0) Bragg peak of the FLL.
Each rocking curve was fitted by a Lorentzian (fig.2) in order to extract the integrated scattered intensity and rocking curve width.

The intensity for a (hk) reflection is 
\begin{equation}
\label{intensity}
I_{hk}=2 \pi V \phi (\gamma /4)^2 \lambda_n^2/\phi_0 \mid F_{hk} \mid^2/Q_{hk}  
\end{equation}
where V is the illuminated sample volume, $\phi$ the incident flux, $\gamma$ the magnetic moment of the neutron, and $\phi_0$ the flux quantum.
 For a quantitative analysis of the scattered intensity at high fields, the Abrikosov limit
can be used in the range $ 0.4.B_{c2}\leq B \leq B_{c2}$ \cite{brandt}, where the Fourier factor of the hexagonal vortex lattice can be expressed as: 

\begin{equation}
\label{fourier}
F_{hk}=(-1)^{\nu} e^{-3^{-1/2} \pi  \nu} \mu_0 M
\end{equation}

where $\nu=(h^2+k^2+hk)$ and $\mu_0.M=(B-B_{c2})/\beta_a(2 \kappa^2-1)$ according to the Abrikosov limit ($\beta_a=1.16$ for the hexagonal lattice, $\kappa$ is the Ginzburg-Landau parameter). 

After normalization to the incident neutrons flux, $F_{10}$ can be extracted from equation \ref{intensity}.
 We choose to fix $B_{c2}=2.55 T$ which was robustly determined, and only one parameter $\kappa$ is adjustable.
 $F_{10}$ compares favorably at high fields with the Abrikosov form of the intensity with $\kappa \approx  12$ (fig.2), giving $\lambda=\kappa .\xi = 136 nm$ with $\xi^2 = \phi_0/(2\pi B_{c2})\approx 128 nm^2 $. To describe the data at the lowest fields, we used 
numerical corrections from Ginzburg-Landau solution as described in \cite{brandt,yaouanc}.
 Since it converges to the Abrikosov limit for $B/B_{c2} \geq  0.4$, there is no supplementary fitting and the agreement with the experimental data is satisfactory (fig.2).
The values of $\lambda$ and $\xi$ are consistent with the literature, considering that previous estimates of $\lambda$ are quite scattered over the broad range $70-250 nm$ (\cite{lambda} and references herein).
The full-width-half-maximum ($FWHM$) of the rocking curves was analyzed with the correction of instrumental resolution, mainly due to 
wavelength spread and the beam divergence. It can be analytically estimated as a first approach using a Gaussian approximation \cite{cubitt}.
 For a peculiar set-up, we find for example $FHWM(resolution)=0.16 deg$ for $B=0.4T$, to compare with $FWHM(rocking-curve)=0.22 deg$, and an intrinsic with $\Delta \omega$ can be extracted \cite{nb}.
Finally, a longitudinal correlation length $d \approx a_0/\pi \Delta \omega$ is deduced. As shown in figure 3, we have found that $d \approx 11 \pm 4 \mu m $ is roughly constant for all field values (the crystal is 200 $\mu m$ thick).
 Note that, since $FWHM$ is close to the instrumental resolution, the value of $d$ has rather the meaning of a lower bound \cite{morten}.
This value is however already far larger than the one deduced from a criteria used to indicate the appearance of dislocations in the FFL
 (the order state is expected to be unstable for $d\leq  20 a_0$ \cite{bragg}
 and we measure $d >  200 a_0$). Due to the poor resolution in the positional correlation function in our SANS geometry,
 we can not distinguish between a perfect lattice of flux lines,
 a Bragg glass state with algebraically decaying translational order \cite{giam2}, or a multi-domains structure which fractures at large scale \cite{gautam}. Detailed investigations of correlation functions of the FLL
 using a high resolution geometry indicate some large scale fracturing of FLL even in best quality $Nb$ samples \cite{tedmc}.
We summarize the most important result of this part: our analysis of FLL Bragg peaks stabilized after ZFC shows that the FLL order is unchanged and robust far inside the peak effect where the critical current has strongly increased.

In $NbSe_2$, FC states show different properties than ZFC states \cite{transport}.
 In particular, they present a large and metastable critical current and are macroscopically heterogeneous,
 even for magnetic fields smaller than $B_{peak}$ \cite{transport,paxy}.
 This is a quite unusual case, since FC leads usually to a more perfect lattice in samples without the peak effect.
 Comparison between FC and ZFC scattering intensity was here tricky.
 Rocking the sample through the expected Bragg angle after FC gives a scattered intensity significantly reduced.
 One might conclude that much of the intensity was incoherently scattered from
 a strongly disordered supercooled state.
However, as noted in \cite{andrew} when measuring $CeRu_2$,
 if the Bragg peaks are strongly broadened, all the intensity is not recovered
 for restricted angular scans. This is also the case if the Bragg angle has changed. Measurements performed with
 large enough angular scans show that most of the intensity in the FC rocking curve
 was surprisingly outside the regular Bragg conditions (fig.4).
 The shape of the rocking curve is difficult to extract because it emerges only slightly from the background.
 However, an analysis with two similar Bragg peaks rotated from the theoretical Bragg angle is most satisfactory.
 Even if two peaks in the rocking curve indicates a very peculiar FLL structure, the sum of integrated intensity of the two peaks corresponds
 to the intensity expected for the regular Abrikosov lattice measured after ZFC (fig.2) \cite{notebene}. The $Q$ value is also unchanged and its radial width is resolution limited.
Since the intensity in the RC is preserved, all the "disorder" should be along the longitudinal
 direction, without affecting the orientational quality of the FLL. This is consistent with measurements
 of the orientational width of the Bragg peaks in the plane of the detector.
 They remain resolution limited ($\Delta \psi =13 \pm 1 deg$ in the azimuthal direction) for the FC case (fig.5). 
Due to this relatively poor azimuthal resolution, we do not know if the FC FLL presents an orientational order as good as the ZFC FLL.
 But clearly, the FC FLL is not
a supercooled glassy-like state which should lead to a degenerated orientational order (a ring of scattering) \cite{ling}.
The double peak structure was first observed in \cite{paxy}. It implies that the FLL is tilted from the magnetic field direction due to magnetic field components in the plane.
 A non symmetric distribution of superficial screening currents was proposed as a qualitative interpretation \cite{paxy}.
 The new informations obtained here (the preserved intensity of the whole FLL after FC, and the absence of bulk transition in the ZFC FLL when crossing the peak) are consistent with this scenario.
  We have tried the usual "shaking" procedure consisting in a variation of the magnetic field in the form of a sinusoidally damped
oscillation, in order to improve the FLL spatial order \cite{levett}. The effect was only marginal, likely due to the non-trivial nature of the FC FLL structure.
 Contrarily, a thin Bragg peak centered at the theoretical Bragg angle is recovered when passing a large dc overcritical current in the sample \cite{paxy}.
 A possible reason of the negligible effect of the "shaking" is that the screening currents are too large to be affected by the currents induced by the field oscillation.
 Another one is that the screening currents have to be polarized in one direction to recover an homogeneous distribution and a regular FLL.
  Note that our results appear consistent with the
 edge contamination model \cite{paltiel} and with the observation of two coexisting macroscopic vortex states \cite{shobo},
 to the extent that the supercooled state with the high critical current stabilizes surface currents only and is unrelated to a bulk transition. Such non homogeneous distribution
 of surface currents is also consistent with the complicated current flow necessary to induce the interface dynamics between the coexisting FLL states \cite{shobo}.
The origin of two competing surface currents is unclear, but may arise from a coexistence of two separate mechanisms contributing to the surface transport current in the mixed state.
 Such a scenario was proposed some years ago from angular magneto-transport measurements coupled with surface
 treatments in conventional type II superconductors \cite{swartz}, and deserves a peculiar attention.
 
In conclusion, we have studied the FLL structure in a $Fe$ doped $NbSe_2$ crystal,which shows a large peak effect in the critical current.
 The ZFC FLL is always ordered. The FC FLL is notably different, but does not present the characteristics of a bulk, glassy-like, disordered state. These results are extremely difficult to conciliate
 with a fracturing or melting transition of the FLL
 usually proposed as the cause of the peak effect and calls for alternative
 interpretations. A direct consequence of our SANS study of FLL structures is that the peak effect may correspond
 to an anomalously large surface or boundary current developing in the peak. It can be frozen over macroscopic zones of the sample and leads to the two states
 mixture responsible for the metastable transport measurements.

\newpage

\begin{figure}
\begin{center}
\includegraphics*[width=8.0cm]{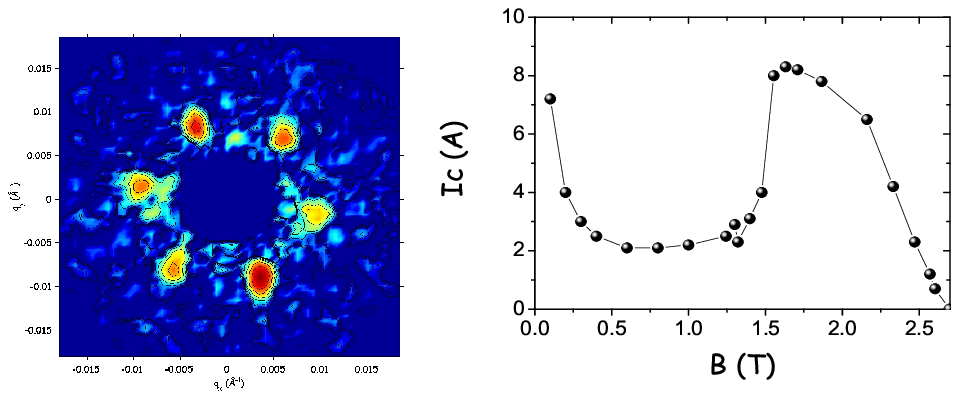}
\end{center}
\caption{Left: 2D pattern of the ZFC FLL ($B=0.4T, T=2K$) obtained by rocking the sample through 
the Bragg conditions and after a subtraction of the normal state background.
 Right: critical current $I_c(B)$ measured after ZFC at $T=2K$ in superfluid Helium and showing the sharp increase of critical current: the peak affect at $B_{peak}\approx 1.5 T$.}
\label{fig.1}
\end{figure}

\begin{figure}
\begin{center}
\includegraphics*[width=8.0cm]{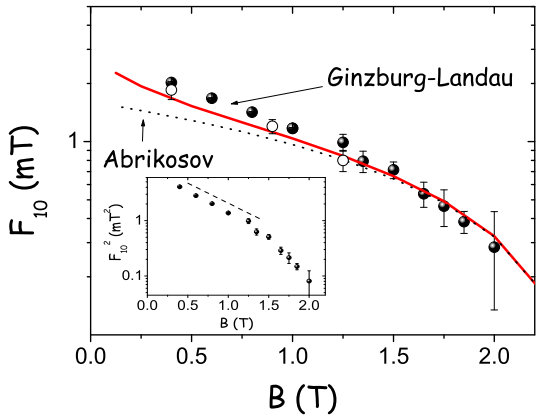}
\end{center}
\caption{The full points are the experimental Fourier component $F_{10}$ deduced from the scattered intensity $I_{10}$ of the ZFC FLL and compared to the Abrikosov and to the Ginzburg-Landau models ($B_{c2}=2.5 T$, $\lambda= 120 nm$).
 The empty points are experimental points corresponding to the FC FLL. In the inset is shown the square of the Fourier component to illustrate the expected cross-over at $B \approx 0.4.B_{c2}=1 T$ from first order Ginzburg-Landau correction to the London model at low field
 (shown as a straight line in this scale) to the Abrikosov line.}
\label{fig.2}
\end{figure}

\begin{figure}
\begin{center}
\includegraphics[width=8.0cm]{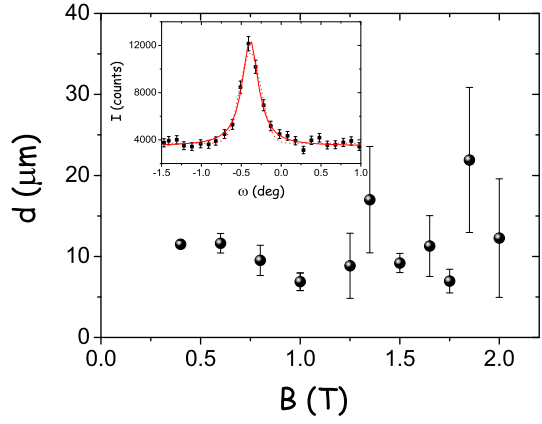}
\end{center}
\caption{Correlation length $d$ deduced from the rocking curve width of a first order 
 Bragg peak as function of the magnetic field ($T=2K$) after ZFC. In the inset is shown a typical rocking
 curve better fitted with a Lorentzian ($B=1.25 T$). No background is subtracted.}
\label{fig.3}
\end{figure}

\begin{figure}
\begin{center}
\includegraphics*[width=8.0cm]{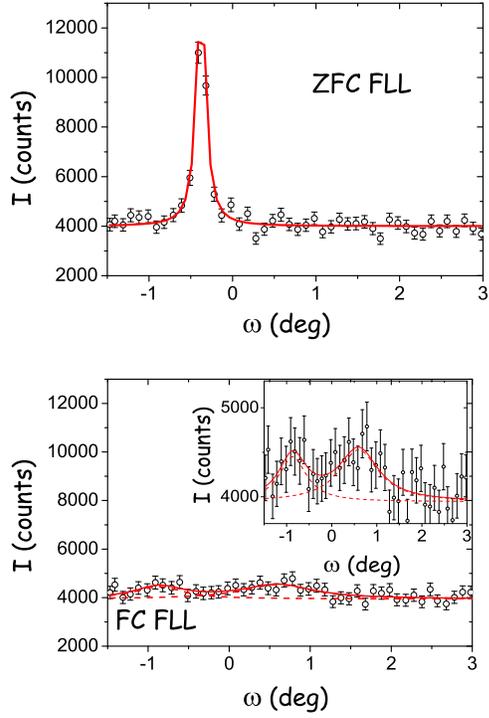}
\end{center}
\caption{Top: rocking curve measured after a zero field cooling for $B=0.4T$.
 The fit is a Lorentzian with $FWHM=0.13 \pm 0.01 deg$. Bottom: rocking curve measured
 after a field cooling for $B=0.4T$ with the same scale as the top figure. The inset
 is a zoom showing two Bragg peaks. The fit is with two Lorentzians of $FWHM_1=0.7 \pm 0.2 deg$ and $FWHM_2=1.0 \pm 0.2 deg$
 which are roughly symmetric with respect to the theoretical Bragg Angle. No background is subtracted. The $Q$ value is the same for FC and ZFC FLL.}
\label{fig.4}
\end{figure}

\begin{figure}
\begin{center}
\includegraphics*[width=8.0cm]{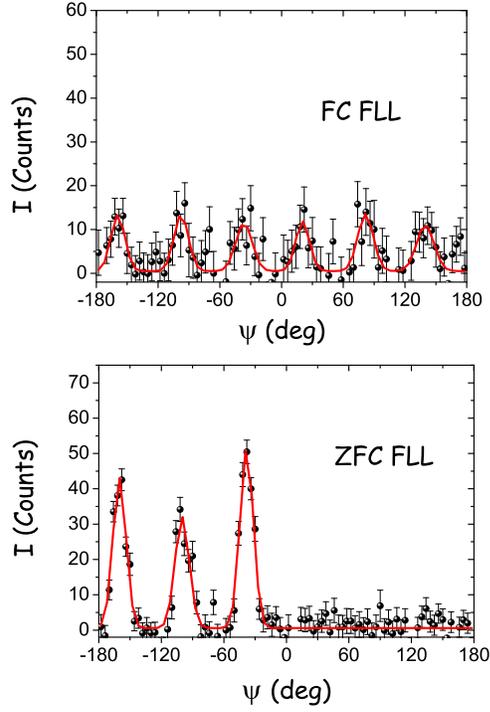}
\end{center}
\caption{Azimuthal scan for FC and ZFC states ($B=0.4T$). Despite $\omega$ rocking curves are very different, 
the orientational order remains resolution limited as demonstrated by the mean value $\Delta \psi$ of the peaks
 FWHM (ZFC:$\Delta \psi =12.98	\pm 0.50 deg$, FC:$14.09\pm	1.38 deg$). For the FC case, all the 6 peaks are observed without
 performing full scans for all the (h,k) reflections because the lattice is in Bragg condition for an extended angular range.}
\label{fig.5}
\end{figure}


\begin{references}
\bibitem{nb} We use a Gaussian approximation to add the resolution contributions (FWHM) in quadrature, whereas the RC are better fitted by Lorentzian, so the correction is an approximation.
\bibitem{notebene} Note that the error bars are large due to the small ratio counts/background (fig.4) (we estimate that $10 \%$ is the typical relative uncertainty).
\end{references}
\end{document}